\documentclass[
showpacs,amsmath,amssymb]{revtex4}
\usepackage{graphicx}
\usepackage{txfonts}
\begin{document}

\title{Exploring holographic dark energy model with Sandage-Loeb test}
\author{Hongbao Zhang\footnote{Current address is Perimeter Institute for Theoretical Physics,
31 Caroline st. N., Waterloo, Ontario N2L 2Y5, Canada.}}
 \affiliation{Department of Astronomy,
Beijing Normal  University, Beijing, 100875, China}
\author{Wuhan Zhong}
\affiliation{Department of
Astronomy, Beijing Normal  University, Beijing, 100875, China}
\author{Zong-Hong Zhu}
\email{zhuzh@bnu.edu.cn}
\affiliation{Department of Astronomy,
Beijing Normal  University, Beijing, 100875, China}
\author{Song He}
\affiliation{Institute of Theoretical Physics, School of Physics,
Peking University, Beijing, 100871, China}

\date{\today}

\begin{abstract}
Taking into account that Sandage-Loeb test is unique in its coverage
of the redshift desert and available in the near future, we explore
the cosmic time evolution behavior of the source redshift for
holographic dark energy model, an important competing cosmological
model. As a result, we find that Sandage-Loeb test can provide a
extremely strong bound on $\Omega^0_m$, while its constraint on
another dimensionless parameter $\lambda$ is weak. In addition, it
is proposed here for the first time that we can also constrain
various cosmological model by measuring the value of $z_{max}$ at
which the peak of redshift velocity occurs. Combining this new
proposed method with the traditional Sandage-Loeb test, we should be
able to provide a better constraint on $\lambda$, at least from the
theoretical perspective.
\end{abstract}

 \maketitle
\section{Introduction}
The recent observations of type Ia Supernovae (SN Ia) indicate that
our universe is currently accelerating\cite{Riess1,Perlmutter}.
Besides the cosmological constant, there have been other various
dark energy models proposed to explain this exotic
phenomenon\cite{Padmanabhan,PR}. On the other hand, a renewed
interest has also been stimulated towards classic cosmological
tests, including the spacial geometry of our universe and the
kinematics of the expansion. For example, the position of acoustic
peaks in the cosmic microwave background (CMB) angular power
spectrum shows the spacial curvature of the
Friedmann-Lema\^{i}tre-Robertson-Walker (FLRW) metric is nearly
flat\cite{Spergel}. A similar test is also carried out by the
detection of baryon acoustic oscillations (BAO) in the power
spectrum of matter calculated from galaxy samples. In addition, the
luminosity distance of SN Ia and other standard candles allows to
provide a constraint on the value of the expansion rate at near
redshifts $z< 2$\cite{Riess2}.

Until now, however, there are still a number of theoretical models
surviving such observational tests. Thus to check the internal
consistency of the underlying cosmological model and discriminate
various competing candidates, some new cosmological tools have been
proposed and performed, such as the lookback time to galaxy
clusters, the age of the universe, and the relative ages of
passively evolving galaxies\cite{Dalal,Rebolo,Capozziello,Simon}. In
particular, recently Corasaniti {\it et al.} employed Sandage-Loeb
test to constrain dark energy models with high significance within
the redshift desert $2<z<5$, where other dark energy probes are
unable to provide useful information about the expansion history of
our universe\cite{Corasaniti}. Later, Balbi and Quercellini extended
this analysis to more general dark energy models\cite{BQ}. But they
all neglected to investigate an important and popular competing
candidate, i.e., holographic dark energy model with Sandage-Loeb
test. Thus as a further step along this line, the purpose of this
paper is to explore the potential constraint on holographic dark
energy model with Sandage-Loeb test.

In the subsequent section, we shall provide a brief review of
holographic dark energy model, including the latest observational
constraints on it. After Sandage-Loeb test is reviewed, we shall
extensively investigate its potential power in constraint on
holographic dark energy model, where we furthermore go beyond the
traditional Sandage-Loeb test within the redshift desert to propose
a new cosmological probe at low redshifts to constrain the model
better. Concluding remarks are presented in the last section. In
addition, as motivated by inflation, the flat FLRW universe is
assumed in the following discussions.
\section{Holographic Dark Energy Model with Its Available Observational Constraints}
Some time ago, taking into account the insightful viewpoint that the
UV cut-off in effective quantum field theory is connected with the
IR cut-off against the formation of black holes, Cohen {\it et al.}
argued that if $\rho_H$ is the zero-point energy density induced by
the UV cut-off, the total energy in a region of size $L$ should not
exceed the mass of black hole of the same size, i.e., $\rho_HL^3\leq
LM^2_p$ with the Planck mass $M_p=\frac{1}{\sqrt{8\pi
G}}$\cite{Cohen}. If we apply this to our universe, then the
zero-point energy can serve as dark energy, referred to as
holographic dark energy. As suggested by Li\cite{Li}, the
corresponding energy density can further be expressed as
\begin{equation}
\rho_H=\frac{3\lambda^2M^2_p}{L^2}.\label{H}
\end{equation}
Here $\lambda$ is a dimensionless parameter, to be determined by the
future complete quantum gravity theory, and $L$ takes the size of
the future event horizon of our universe, i.e.,
\begin{equation}
L[a(t)]=a(t)\int_t^\infty\frac{dt'}{a(t')}=a(t)\int_{a(t)}^\infty\frac{da'}{H'a'^2},\label{F}
\end{equation}
where $a$ is the scale factor of our universe, $t$ is the cosmic
time and $H$ is Hubble parameter.

Next let us consider a flat FLRW universe with a matter component
$\rho_m$ (including both baryon matter and cold dark matter) and a
holographic dark energy $\rho_H$. Then Friedmann equation reads
\begin{equation}
3M^2_pH^2=\rho_m+\rho_H,
\end{equation}
which can also be expressed equivalently as
\begin{equation}
\frac{H^2}{H^2_0}=\frac{\Omega^0_m}{a^3}+\Omega_H\frac{H^2}{H^2_0}.\label{Friedmann}
\end{equation}
Here $\Omega_m=\frac{\rho_m}{3M^2_pH^2}$ and
$\Omega_H=\frac{\rho_H}{3M^2_pH^2}$. In addition, the superscript
(subscript) $0$ denotes the value for the corresponding physical
quantity at the present time. Especially, for convenience but
without loss of generality, we have set the present scale factor
$a_0=1$ here. Later combining Eq.(\ref{H}) with Eq.(\ref{F}), we
have
\begin{equation}
\int_a^\infty\frac{d\ln
a'}{H'a'}=\frac{\lambda}{Ha\sqrt{\Omega_H}}.\label{X}
\end{equation}
On the other hand, Eq.(\ref{Friedmann}) gives
\begin{equation}
\frac{1}{Ha}=\frac{\sqrt{a(1-\Omega_H)}}{H_0\sqrt{\Omega^0_m}}.\label{Y}
\end{equation}
Substituting Eq.(\ref{Y}) into Eq.(\ref{X}) and taking derivative
with respect to $\ln a$ on both sides, we obtain
\begin{equation}
\frac{d\Omega_H}{d\ln
a}=\Omega_H(1-\Omega_H)(1+\frac{2}{\lambda}\sqrt{\Omega_H}),\label{D}
\end{equation}
which describes the dynamic evolution of holographic dark energy,
and can be formulated in terms of the redshift of our universe
$z=\frac{1}{a}-1$ as
\begin{equation}
\frac{d\Omega_H}{\Omega_H(1-\Omega_H)(1+\frac{2}{\lambda}\sqrt{\Omega_H})}=-\frac{dz}{1+z}.\label{Z1}
\end{equation}
It can further be integrated analytically as follows
\begin{equation}
F_\lambda(\Omega_H)\equiv\ln\Omega_H-\frac{\lambda}{2+\lambda}\ln(1-\sqrt{\Omega_H})+\frac{\lambda}{2-\lambda}\ln(1+\sqrt{\Omega_H})-\frac{8}{4-\lambda^2}
\ln(\lambda+2\sqrt{\Omega_H})=-\ln(1+z)+F_\lambda(1-\Omega^0_m).
\end{equation}
If we write the equation of state as $w_H=\frac{p_H}{\rho_H}$ for
holographic dark energy, then the conservation of energy momentum
gives
\begin{equation}
\dot{\rho}_H+3H(1+w_H)\rho_H=0,
\end{equation}
which implies
\begin{equation}
w_H=-1-\frac{1}{3}\frac{d\ln\rho_H}{d\ln
a}=-\frac{1}{3}(1+\frac{2}{\lambda}\sqrt{\Omega_H}),
\end{equation}
where we have employed
$\rho_H=\frac{\Omega_H\rho^0_m}{(1-\Omega_H)a^3}$ and Eq.(\ref{D})
in the second step.

Note that there are only three free parameters in this holographic
dark energy model: One is the kinematic parameter $H_0$, whose value
is taken as $72\frac{km}{s\cdot Mpc}$ in the following calculations.
The others are the dynamic parameters $\Omega^0_m$ and $\lambda$,
which determine the large scale evolution of our universe, including
the final fate of our universe. Thus constraint of $\Omega^0_m$ and
$\lambda$ from observational data plays a fundamental role in
diagnosis of this model\cite{footnote}.

Such a dynamical dark energy model has been constrained by various
astronomical observation data, such as the luminosity distance of SN
Ia\cite{HG}, X-ray gas mass fraction of galaxy clusters\cite{Chang},
the relative ages of galaxies\cite{YZ}. In addition, a joint
constraint from SN Ia, CMB and LSS observational data has also been
performed\cite{ZW1,ZW2}. In particular, combining the latest
observational data from SN Ia, and CMB plus LSS, Zhang and Wu
obtained $\Omega^0_m=0.29\pm0.03$ and the dimensionless parameter
$\lambda=0.91^{+0.26}_{-0.18}$ at the $1\sigma$ confidence
level\cite{ZW2}. In the next section, we shall focus ourselves on
constraining holographic dark energy model with Sandage-Loeb test.
\section{Sandage-Loeb Test and Its Potential Constraint on Holographic Dark Energy Model}
It is useful to firstly derive the redshift variation underlying
Sandage-Loeb test. Consider a given source without peculiar
velocity, which emitted its light at a time $t_s$, then the observed
redshift at time $t_0$ is
\begin{equation}
z(t_0)=\frac{a(t_0)}{a(t_s)}-1,
\end{equation}
which becomes after a time interval $\delta t_0$
\begin{equation}
z(t_0+\delta t_0)=\frac{a(t_0+\delta t_0)}{a(t_s+\delta t_s)}-1.
\end{equation}
Then in the linear approximation, the observed redshift variation of
the fixed source gives
\begin{eqnarray}
\delta z(t_0)&\equiv& z(t_0+\delta t_0)-z(t_0)=\frac{a(t_0+\delta
t_0)}{a(t_s+\delta t_s)}-\frac{a(t_0)}{a(t_s)}
\approx\frac{a(t_0)[1+H(t_0)\delta t_0]}{a(t_s)[1+H(t_s)\delta t_s]}-\frac{a(t_0)}{a(t_s)}\nonumber\\
&\approx&\frac{a(t_0)}{a(t_s)}[1+H(t_0)\delta t_0 -H(t_s)\delta
t_s]-\frac{a(t_0)}{a(t_s)}\approx H(t_0)\delta
t_0[1+z(t_0)-\frac{H(t_s)}{H(t_0)}],
\end{eqnarray}
where $\delta t_1\approx[1+z(t_1)]\delta t_s$ has been used in the
last step. This redshift variation can also be expressed in terms of
a spectroscopic velocity shift, i.e.,
\begin{equation}
\delta v=\frac{\delta z}{1+z}=H_0\delta
t_0[1-\frac{E(z)}{1+z}],\label{R}
\end{equation}
where $E(z)=\frac{H(z)}{H_0}$. Note that the redshift variation is
directly related to the expansion rate of our universe, which is the
most essential part of any model. For example, in holographic dark
energy model considered here, by Eq.(\ref{Y}) the expansion rate
reads
\begin{equation}
E(z)=\sqrt{\frac{\Omega^0_m(1+z)^3}{1-\Omega_H}},\label{Z2}
\end{equation}
which is obviously different from the $\Lambda$CDM model with
\begin{equation}
E(z)=\sqrt{\Omega^0_m(1+z)^3+1-\Omega^0_m}.\label{Z3}
\end{equation}
Therefore different from those classical cosmological probes, which
are almost exclusively sensitive to the cosmological parameters
through a time integral of Hubble parameter, the measurement of
velocity shift plays a critical role in investigating the physical
mechanism responsible for the acceleration and discriminating
various dark energy models.

This kind of astronomical observation as a possible cosmological
tool, referred to as Sandage-Loeb test, was firstly put forward by
Sandage\cite{Sandage}, and revisited by Loeb more
recently\cite{Loeb}. With the foreseen development of very large
telescopes, and the availability of spectrographs of unprecedented
resolution, the quasar absorption lines typical of the
Lyman-$\alpha$ forest provide a powerful tool to measure the
velocity shift within the redshift desert\cite{Corasaniti,Loeb}.
Especially, invoking Monte Carlo simulations, Pasquini {\it et al.}
estimated the statistical error on $\delta v$ as measured by the
cosmic dynamics experiment (CODEX) spectrograph over a period of 10
years as
\begin{equation}
\sigma_{\delta
v}=1.4(\frac{2350}{s/n})\sqrt{\frac{30}{N}}(\frac{5}{1+z})^{1.8}\frac{cm}{s},\label{S}
\end{equation}
where $s/n$ denotes the signal to noise ratio per 0.0125A pixel, and
$N$ is the number of Lyman-$\alpha$ quasars\cite{Pasquini}. In what
follows, we assume that the future experimental configuration and
uncertainties is similar to those expected from CODEX. Namely, the
error bars are estimated from Eq.(\ref{S}), with the assumption that
a total of 240 quasars can be observed uniformly distributed in 6
equally spaced redshift bins within the redshift desert, with $s/n =
3000$\cite{Corasaniti,BQ}.

\begin{figure}
  \includegraphics[width=2.4inch]{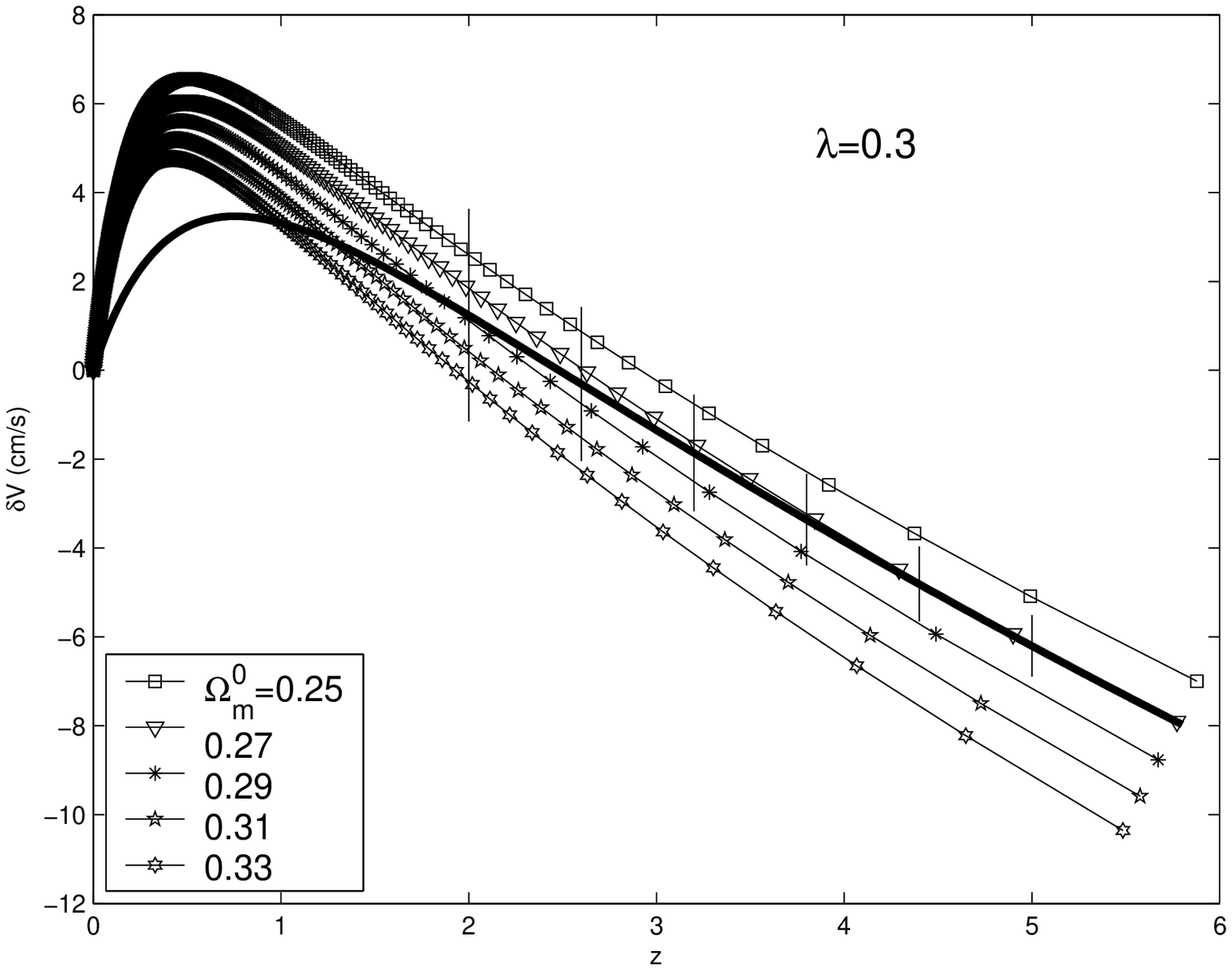}\\
  \includegraphics[width=2.4inch]{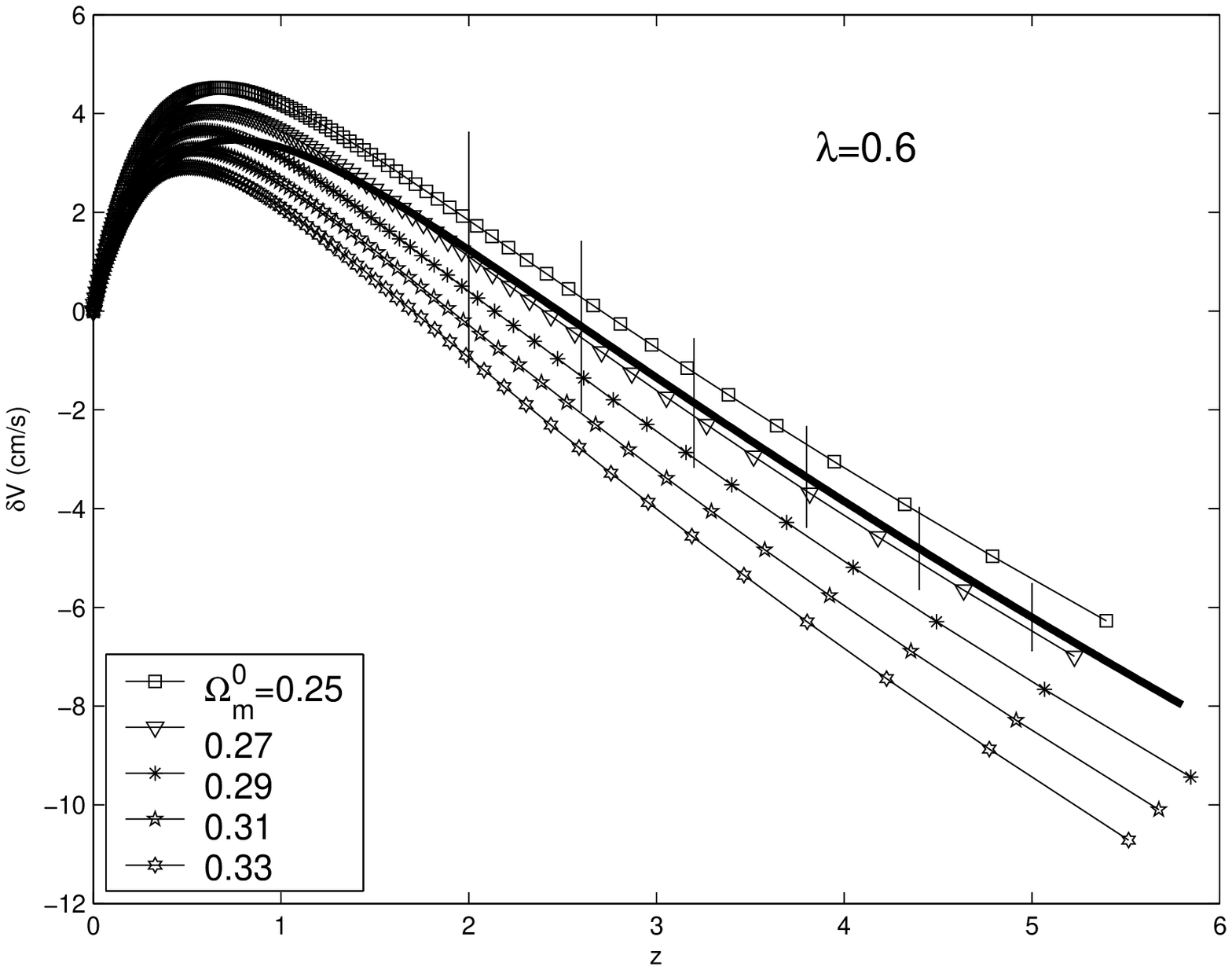}\\
    \includegraphics[width=2.4inch]{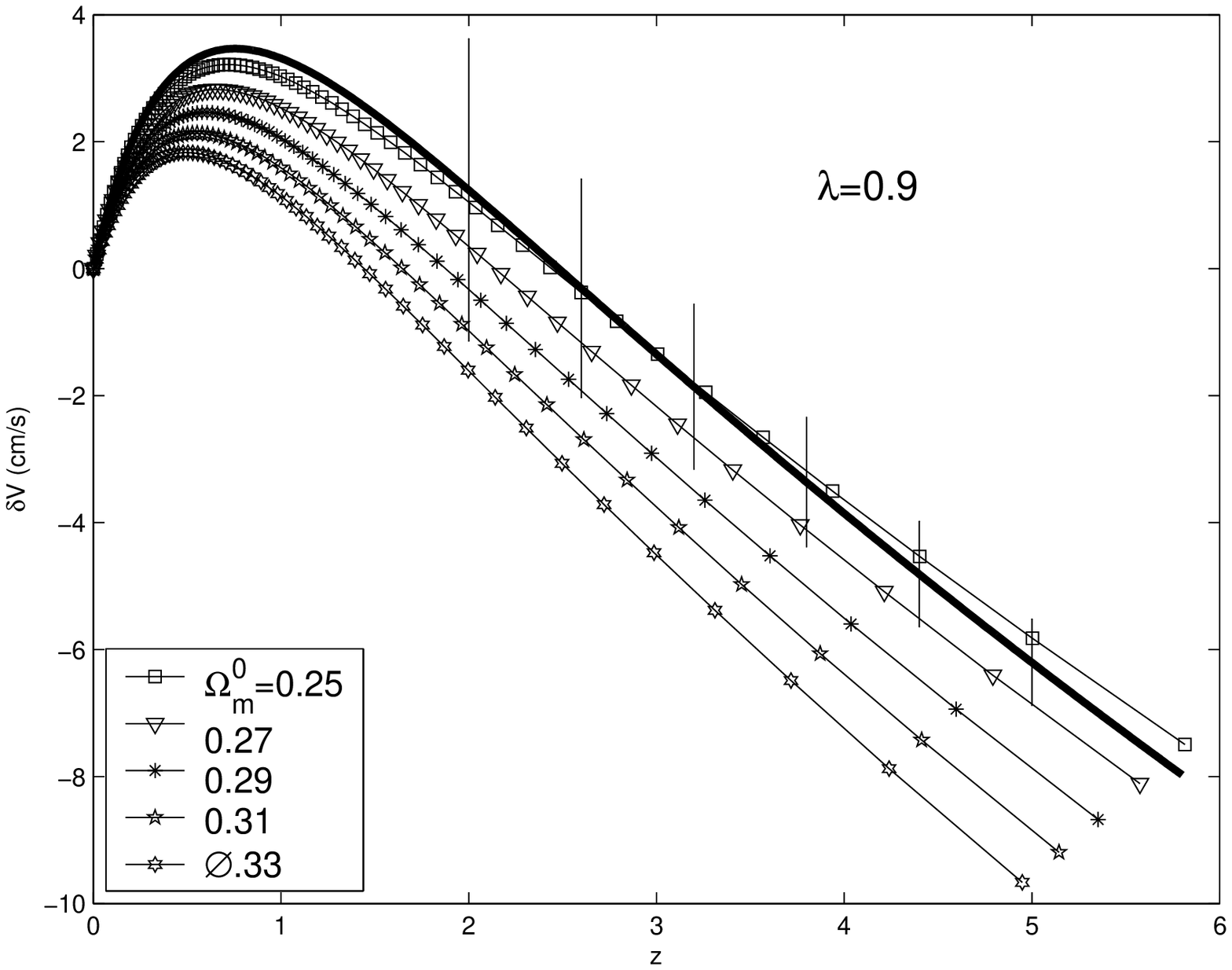}\\
      \includegraphics[width=2.4inch]{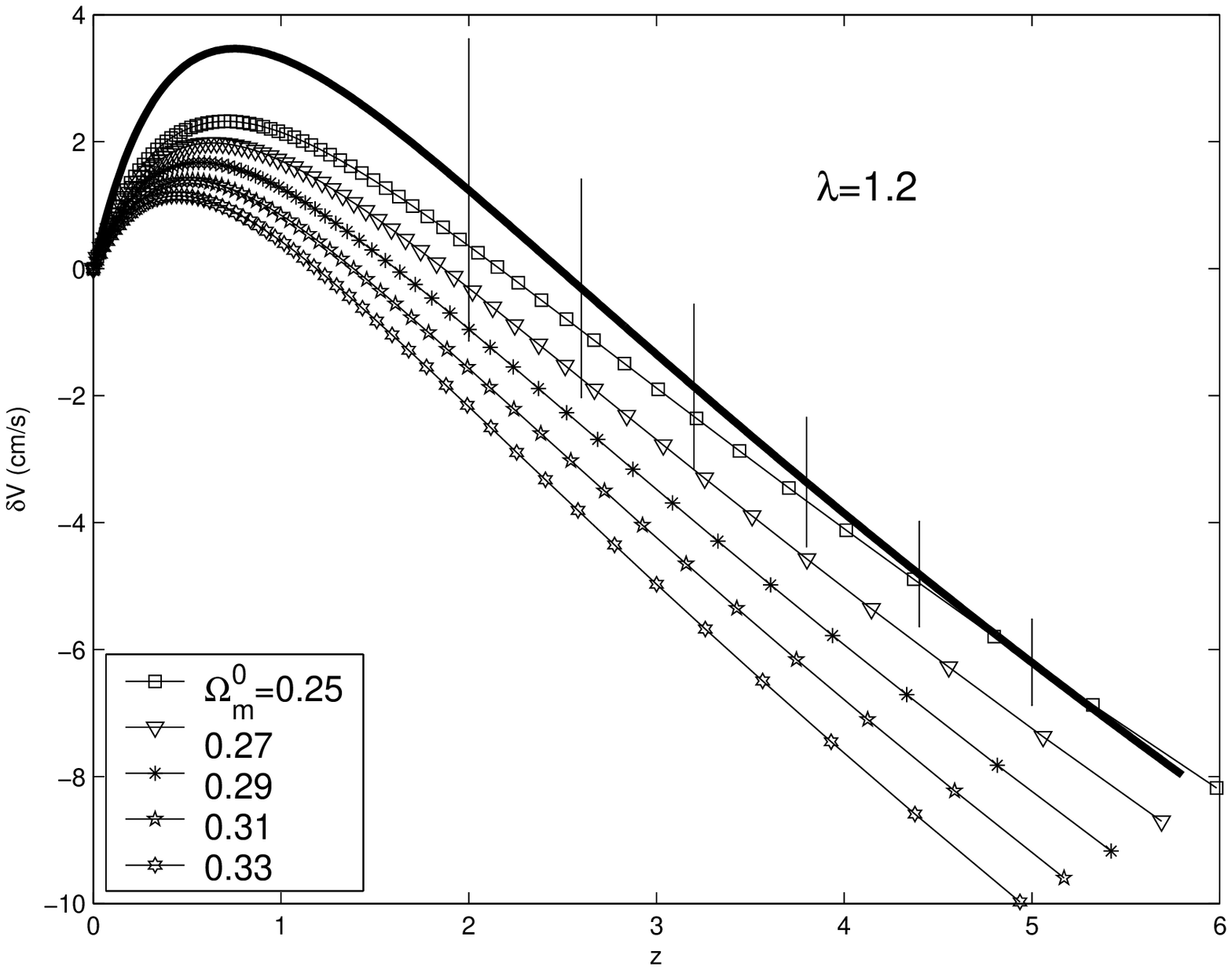}\\
        \includegraphics[width=2.4inch]{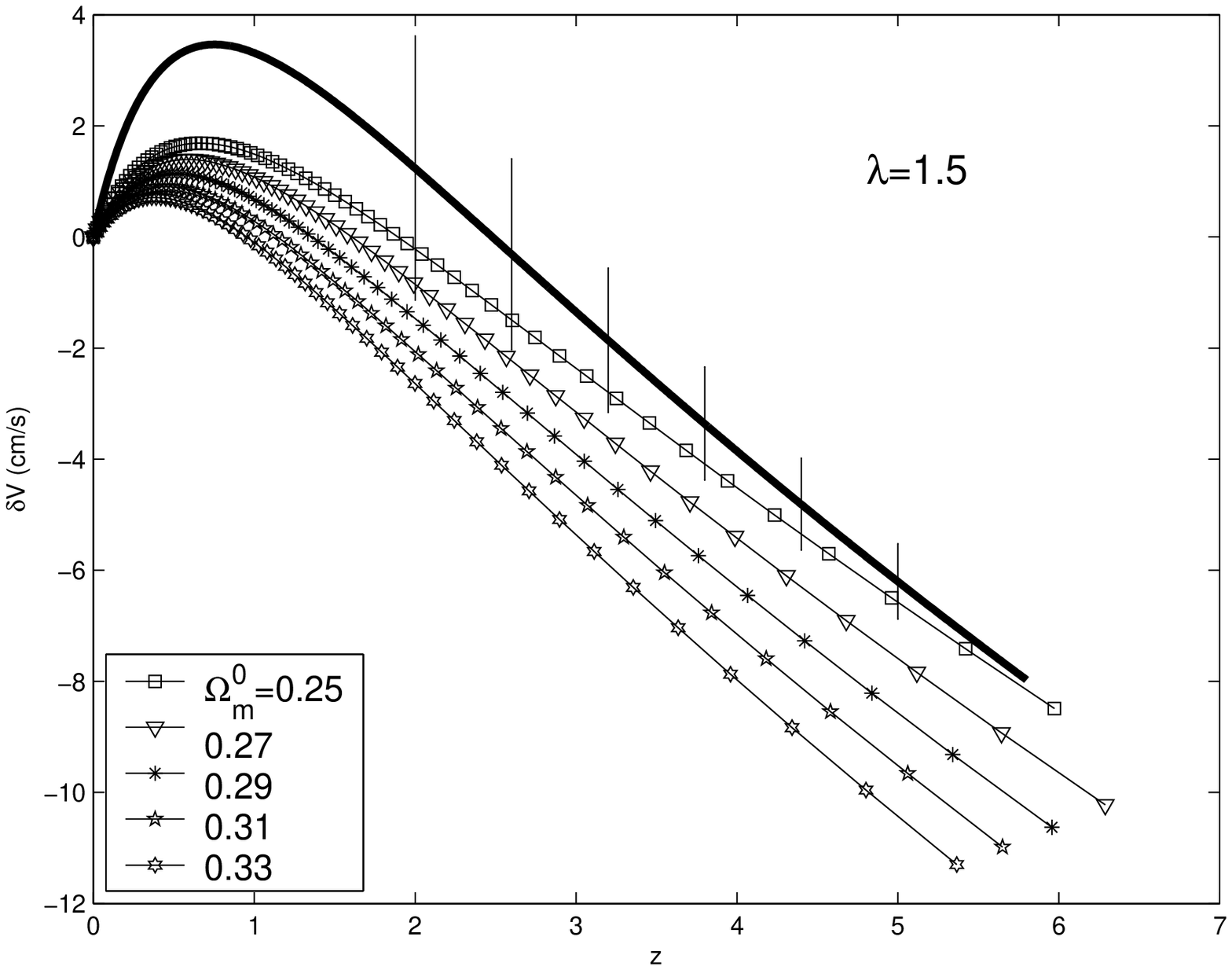}\\
         \caption{The redshift velocity as function of the source redshift
  for fixed $\lambda$ and different values of $\Omega^0_m$.}\label{f1}
        \end{figure}

\begin{figure}
          \includegraphics[width=2.4inch]{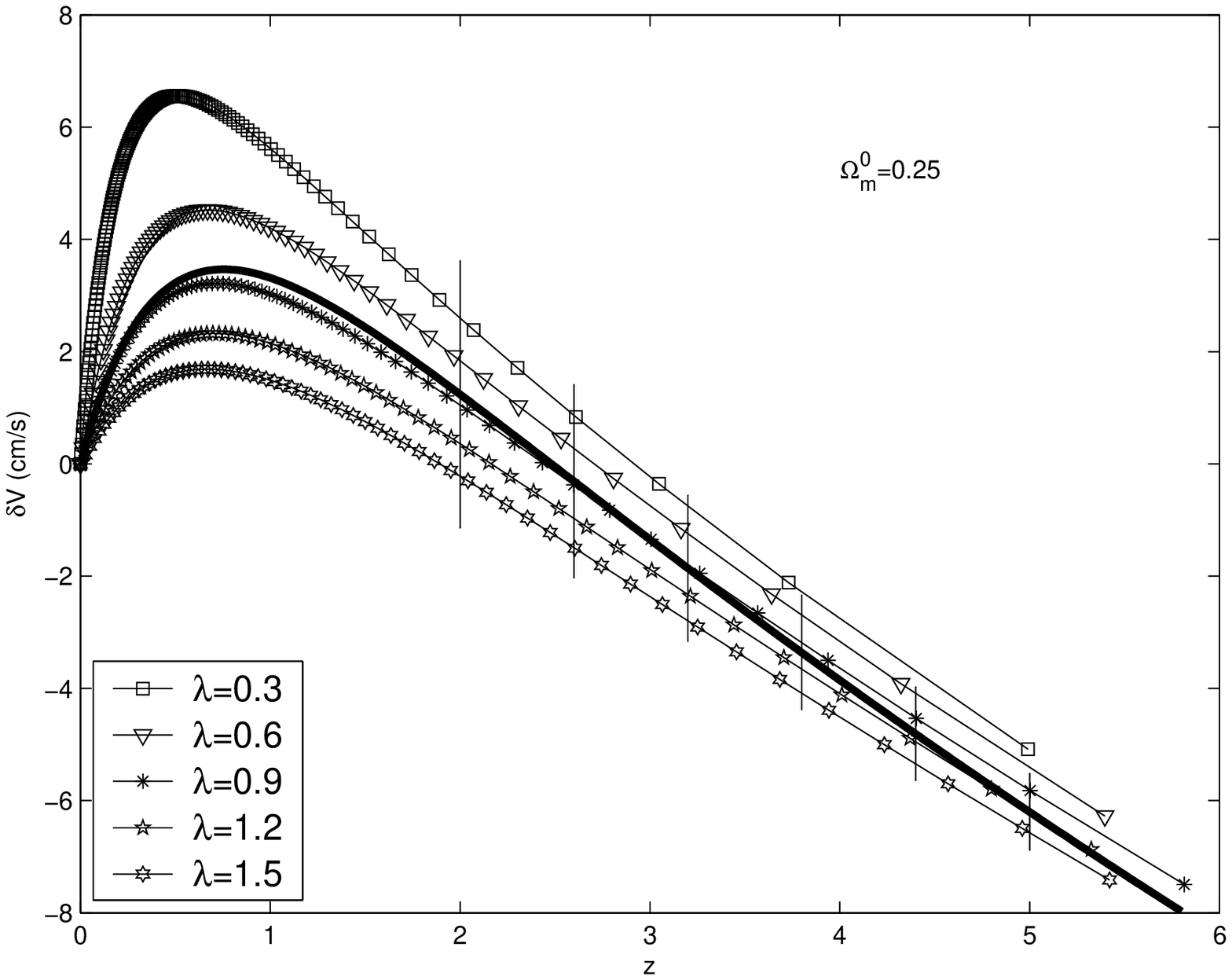}\\
            \includegraphics[width=2.4inch]{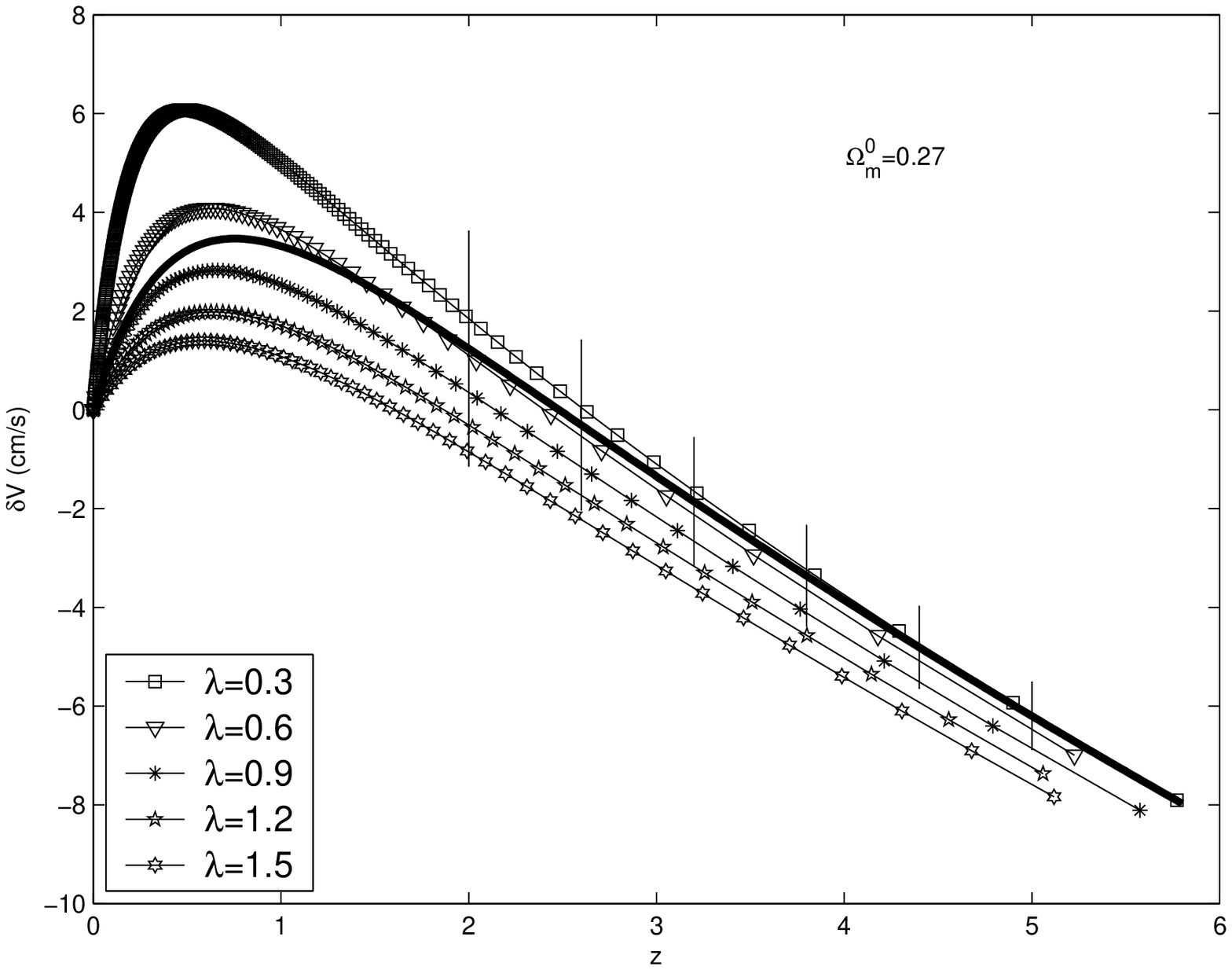}\\
              \includegraphics[width=2.4inch]{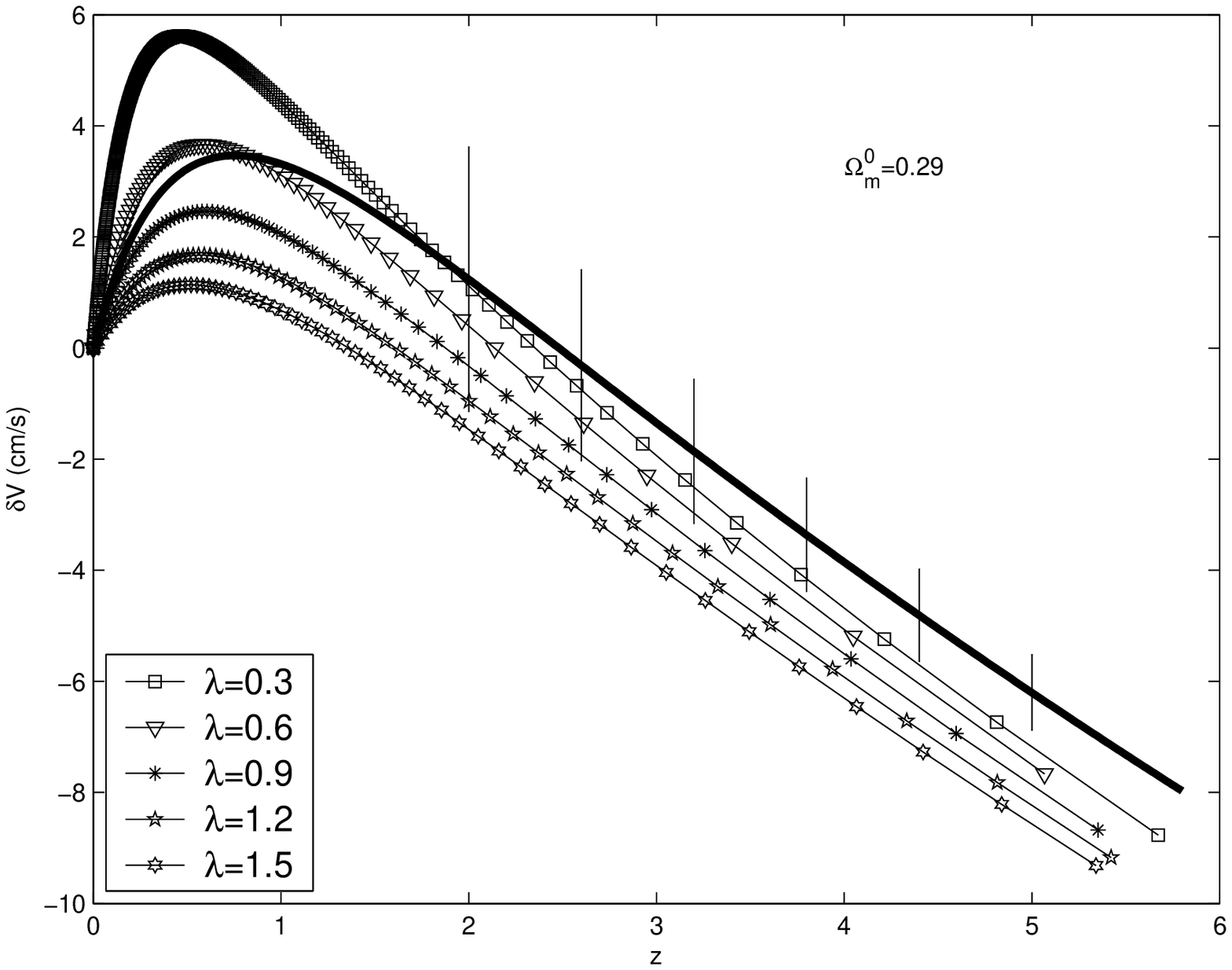}\\
                \includegraphics[width=2.4inch]{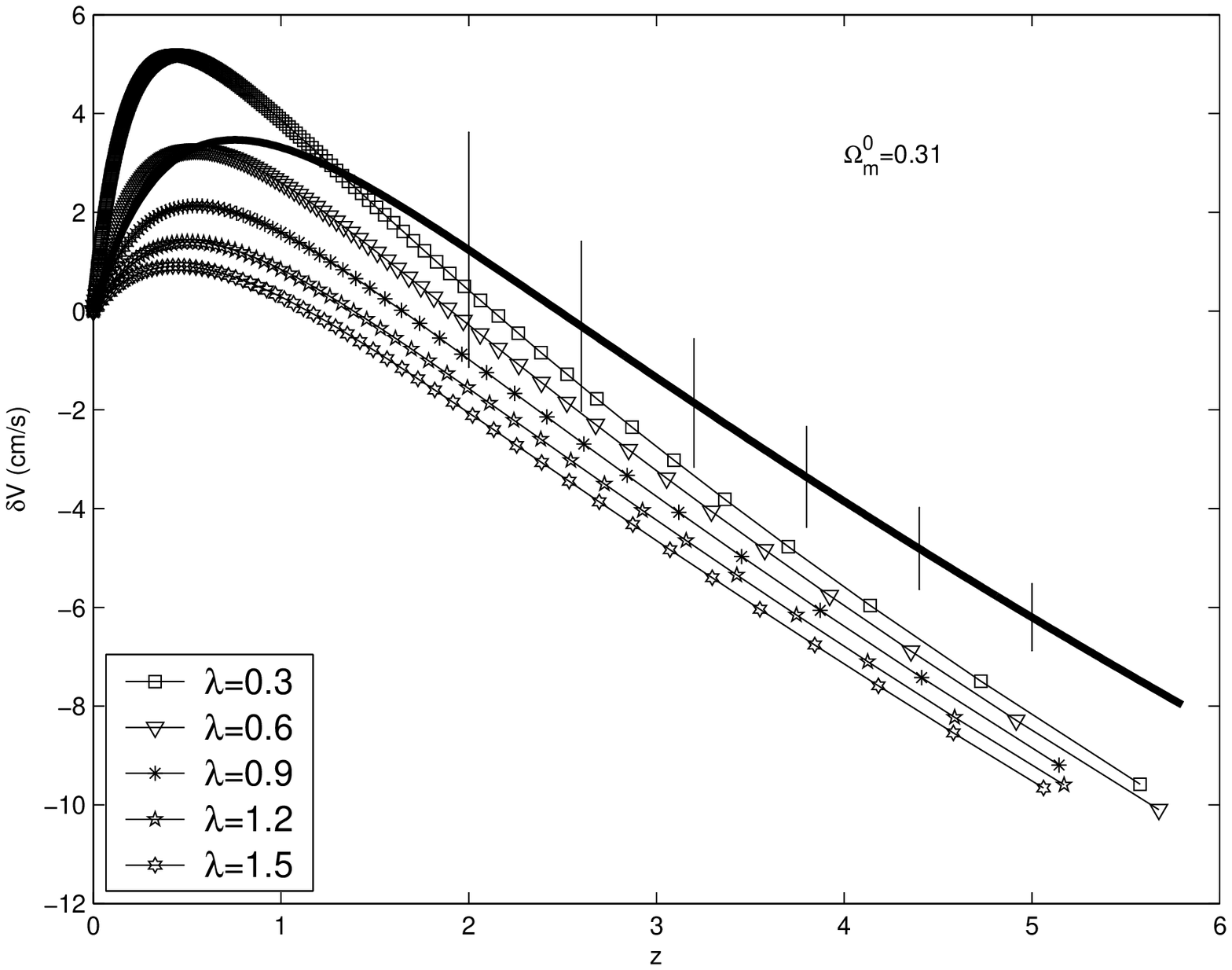}\\
                  \includegraphics[width=2.4inch]{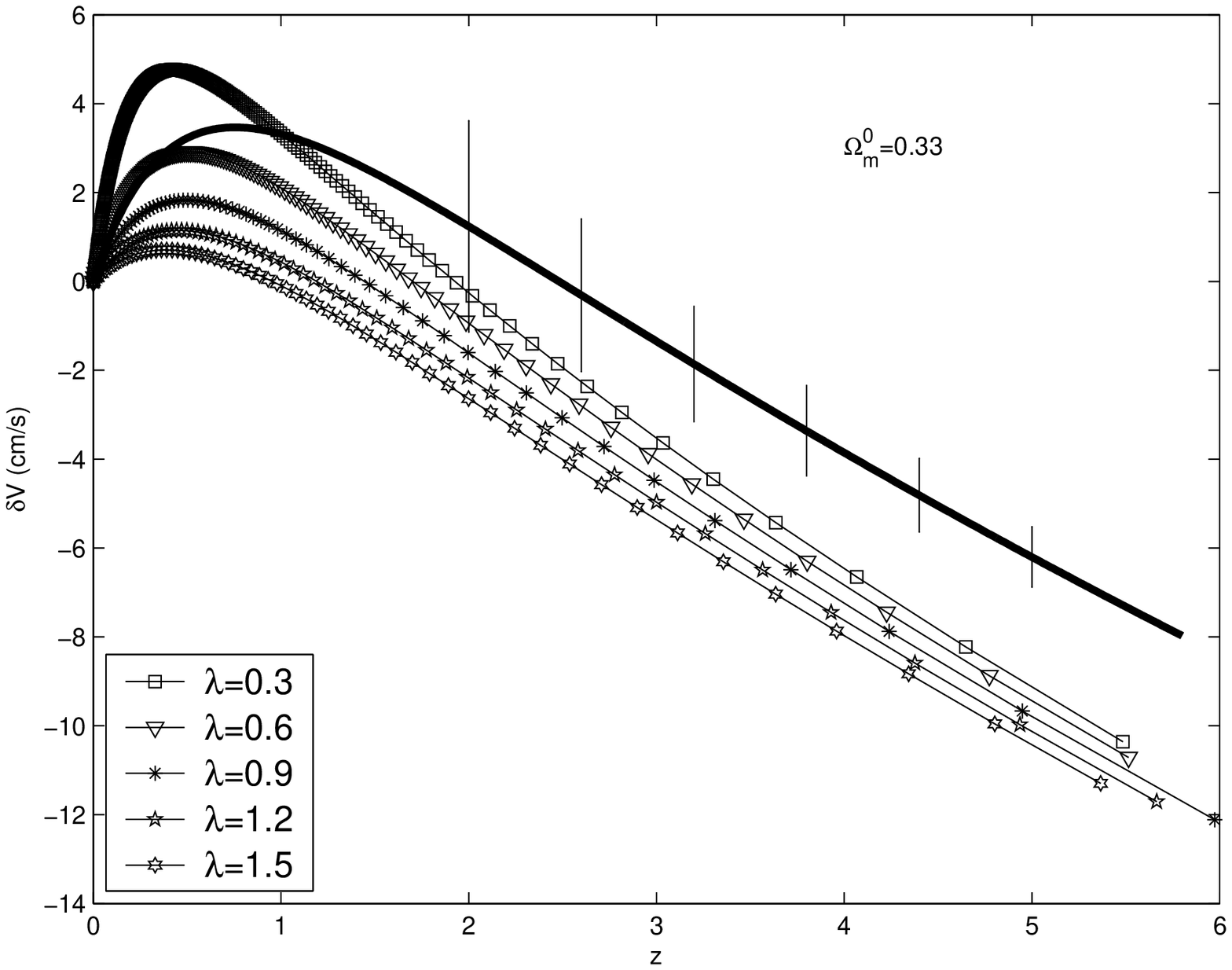}\\
  \caption{The redshift velocity as function of the source redshift
  for fixed $\Omega^0_m$ and different values of $\lambda$.}\label{f2}
\end{figure}

Now let us start to explore the behavior of redshift velocity in
holographic dark energy model. To proceed, we firstly plug
Eq.(\ref{Z2}) into Eq.(\ref{R}), and then perform a numerical
calculation for different values of $\Omega^0_m$ and $\lambda$. The
results obtained are plotted in FIG.\ref{f1} and FIG.\ref{f2}. We
also plot the redshift velocity curve for $\Lambda$CDM model with
$\Omega^0_m=0.27$ with the error bars from Eq.(\ref{S}) in the
figures, which is convenient for us to compare holographic dark
energy model with the fiducial concordance cosmological model.

\begin{figure}
  \includegraphics[width=4.4inch]{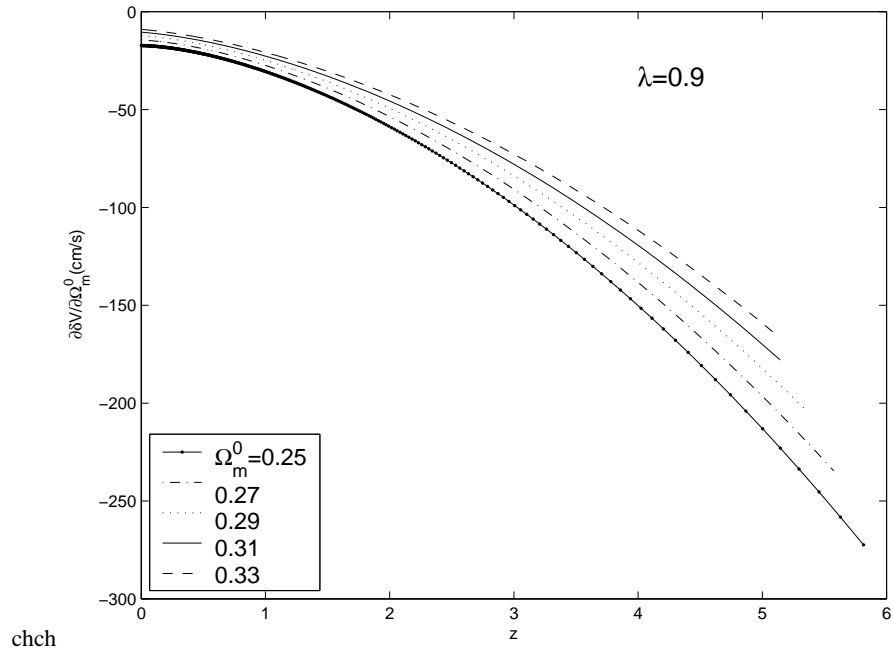}\\
\caption{The derivative of redshift velocity with respect to
$\Omega^0_m$  as function of the source redshift
  for fixed $\lambda=0.9$ and different values of $\Omega^0_m$.}\label{f3}
        \end{figure}

As shown in FIG.\ref{f1}, for fixed $\lambda$,  the differences of
redshift velocity among different values of $\Omega^0_m$ become
bigger and bigger with the increase of the source redshift, which is
also explicitly supported by FIG.\ref{f3}. On the contrary, the
error bars from Eq.(\ref{S}) is a decreasing function of the
redshift. Thus it is advantageous to employ Sandage-Loeb test to
distinguish holographic dark energy models among different values of
$\Omega^0_m$ within the redshift desert. On the other hand,
FIG.\ref{f2} demonstrates that the differences of redshift velocity
decrease with the source redshift for fixed $\Omega^0_m$ but
different values of $\lambda$; furthermore the magnitude of
differences is also comparable to that of error bars. It means that
Sandage-Loeb test is not very sensitive to the dimensionless
parameter $\lambda$. That is, it is difficult to discriminate
holographic dark energy models with different $\lambda$s by
Sandage-Loeb test within the redshift desert.

\begin{figure}

        \begin{center}
  \includegraphics[width=3inch]{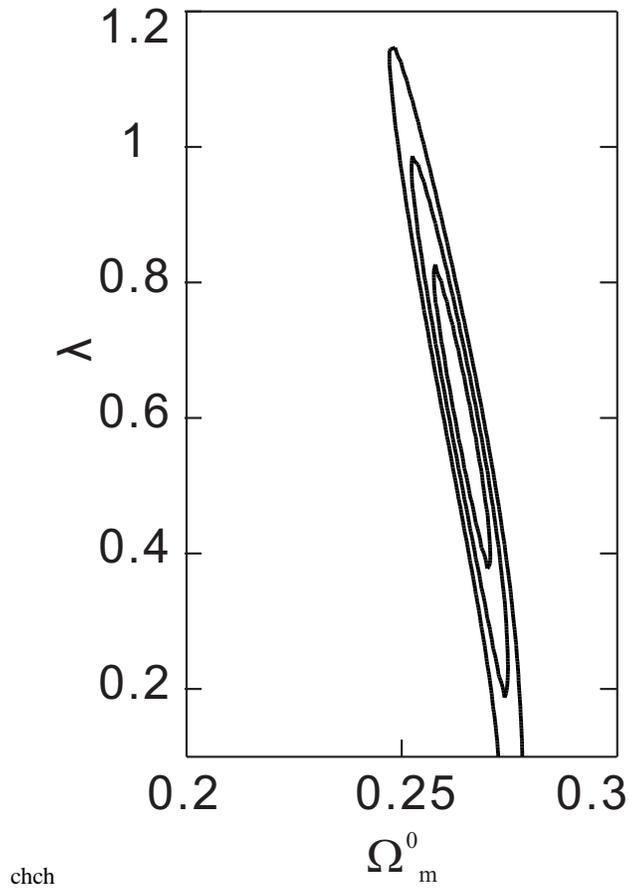}
\caption{Sandage-Loeb test contours for $1\sigma$, $2\sigma$, and
$3\sigma$ respectively.}\label{f4}
\end{center}
        \end{figure}

In addition, if we assume the the prediction of the fiducial
$\Lambda$CDM model with the error bars from Eq.(\ref{S}) represents
the future practical measurement result of the redshift velocity
within the redshift desert, FIG.\ref{f1} and FIG.\ref{f2} show that
Sandage-Loeb test seems to favor small $\Omega^0_m$s, such as
$(\Omega^0_m=0.25,\lambda=0.9,1.2,1.5)$ and
$(\Omega^0_m=0.27,\lambda=0.3,0.6,0.9)$. In order to check this
naive observation, next we would like to perform $\chi^2$ statistics
for the model parameters ($\Omega^0_m$,$\lambda$). With the
assumption considered above, we have
\begin{equation}
\chi^2_{SL}=\sum_{i=1}^{240}\frac{[\delta v_H(z_i)-\delta
v_L(z_i)]^2}{\sigma^2_{\delta v}(z_i)}.
\end{equation}
Here $\delta v_H(z_i)$ and $\delta v_L(z_i)$ represent the
prediction value from holographic dark energy model and the fiducial
concordance model, respectively. In addition, $\sigma_{\delta
v}(z_i)$ is estimated from Eq.(\ref{S}). Accordingly numerical
computation gives the contour diagrams as FIG.\ref{f4}. Especially,
the 1$\sigma$ fit value for the model
parameters:$\Omega^0_m=0.264^{+0.007}_{-0.006}$, and
$\lambda=0.611^{+0.215}_{-0.233}$ with $\chi^2_{min}=0.086$, which
also confirms the aforementioned observation that Sandage-Loeb test
is very sensitive to $\Omega^0_m$, while the constraint on $\lambda$
is weaker.

 Last
but definitely not least, if we do not restrict ourselves within the
redshift desert, it is noteworthy that there is something
interesting appearing in FIG.\ref{f1} and FIG.\ref{f2}: In either
$\Lambda$CDM model or holographic dark energy model, there always
exists some low redshift $z_{max}$ at which the redshift velocity
reaches its maximum\cite{notefoot}. Obviously $z_{max}$ can be
obtained by requiring the usual conditions satisfied, i.e.,
$\frac{d\delta v}{dz}|_{z_{max}}=0$.
We have then from Eq.(\ref{R})
\begin{equation}
\frac{dE}{dz}|_{z_{max}}=\frac{E}{1+z}|_{z_{max}}.\label{P}
\end{equation}
Hereby we find
\begin{equation}
z_{max}=\sqrt[3]{\frac{2(1-\Omega^0_m)}{\Omega^0_m}}-1\label{new1}
\end{equation}
for $\Lambda$CDM model. Similarly, an implicit but analytic formula
can be obtained for $z_{max}$ in holographic dark energy model,
i.e.,
\begin{equation}
\Omega_H(1+\frac{2}{\lambda}\sqrt{\Omega_H})|_{z_{max}}=1.\label{new2}
\end{equation}
Note that $z_{max}$ is related to the dynamic cosmological
parameter(s) in such a direct way, but independent of $H_0$. For
example, if the fiducial concordance model is really correct, we
should find observationally $z_{max}=0.755$ by taking
$\Omega^0_m=0.27$ in Eq.(\ref{new1}). It is thus suggested that the
measurement of $z_{max}$ may provide another strong potential test
of various cosmological models, at least from the theoretical
perspective. Different from the traditional Sandage-Loeb test, where
precise measurement of amplitude of redshift velocity is needed
within the redshift desert, this new possible test need only to
determine $z_{max}$ by discerning a narrow low redshift region of
$z\leq1$ where the peak of redshift velocity occurs, regardless of
the specific value of amplitude of redshift velocity, including the
magnitude of peak. For holographic dark energy model considered
here, if the value of $z_{max}$ can be measured precisely, it is
obvious that we can employ Eq.(\ref{new2}) to provide a stronger
constraint on the dimensionless parameter $\lambda$, in combination
with the traditional Sandage-Loeb test.


\section{Concluding Remarks}
We have explored holographic dark energy model with Sandage-Loeb
test. The obtained result shows that Sandage-Loeb test from the
redshift desert can impose a strong bound on $\Omega^0_m$, while its
constraint on $\lambda$ is weaker. Especially, if we fit holographic
dark energy model to the fiducial $\Lambda$CDM model, which is
assumed to provide a prediction of future measurement value with the
error estimated from Monte Carlo stimulations, we find
$\Omega^0_m=0.264^{+0.007}_{-0.006}$, and
$\lambda=0.611^{+0.215}_{-0.233}$ with $\chi^2_{min}=0.086$ at
$1\sigma$ accuracy level. In addition, we notice an interesting and
significant behavior for the redshift velocity function, i.e., the
peak of redshift velocity seems to always occur at some low
redshift, which may be employed to provide another strong potential
test of various cosmological models. A more detailed analysis of
this new suggested cosmological tool, such as its prospects of
observational availability and feasible constraints on various
cosmological models is worthy of further investigation but beyond
the scope of this paper. We expect to report it elsewhere.

\acknowledgments{We would like to give much gratitude to Dragan
Huterer for his helpful discussion on Sandage-Loeb test and related
issues. Valuable suggestions from Tongjie Zhang are also
appreciated. We also acknowledge Li Chen for her kind help with
numerical calculations. The work by HZ and ZZ was supported 
by NSFC under Grant No.~10533010, 973 Program No.~2007CB815401, Program for New Century
Excellent Talents in University (NCET) of China and the Project-sponsored by SRF
for ROCS, SEM of China. WZ was supported in part by NSFC
under Grant Nos.~10173024 and ~10433030. SH was supported by NSFC
under Grant Nos.~10235040 and ~10421003.}


\begin{thebibliography}{xx}
\bibitem{footnote}Since holographic dark energy is
believed to originate from the quantum gravity effect, constraint of
the dimensionless parameter $\lambda$ should also acquire its
importance in the future complete quantum gravity.
\end{thebibliography}
\end{document}